\documentclass[twocolumn,showpacs,pra]{revtex4}

\usepackage[colorlinks=true,citecolor=blue,linkcolor=magenta]{hyperref}

\usepackage{amsmath}
\usepackage{amssymb}
\usepackage{txfonts}
\usepackage{graphicx}
\usepackage{epstopdf}
\usepackage{dcolumn}
\usepackage{bm}
\begin{document}
\title{Switchable dynamics in the deep-strong-coupling regime}
\author{Li-Li Zheng}
\author{Qian Bin}
\email{binqian94@163.com}
\author{Zhi-Ming Zhan}
\author{Sha Li}
\author{Xin-You L\"{u}}
\email{xinyoulu@hust.edu.cn}
\author{Ying Wu}
\email{yingwu2@126.com}
\affiliation{School of Physics, Huazhong University of Science and Technology, Wuhan, 430074, P. R. China}
\date{\today}
\begin{abstract}
We investigate theoretically the dynamics of the system that consists of a cascade three-level emitter interacting with a single-mode resonator in the deep-strong-coupling regime. We show that the dynamical evolution of the system can only occur in a certain parity chain decided by the initial state, in which the photon population and the initial state probability present periodic collapses and revivals. In particular, we find that the evolution of the dynamics can be controlled by feeding the time-control pulses into the system. Control pluses with specific arrival times can suddenly switch off and on the time evolutions of the system populations and initial state probability when the system is originally in a symmetry superposition state. Physically, the switch-off of the evolution originates from the symmetry-breaking of the state, i.e, $(|g0\rangle+|f0\rangle)/\sqrt{2}\rightarrow(|g0\rangle-|f0\rangle)/\sqrt{2}$. This work offers an all-optical approach to manipulate the dynamics of the system, which might have potential application in modern quantum technology.
\end{abstract}
\maketitle
\section{introduction}
The investigation of light-matter interaction has been one of the central topics of quantum optics for last few decades~\cite{ref1}. Studies of light-matter interaction from the weak-coupling to the strong-coupling regime have been done in a variety of cavity quantum electrodynamics (QED) systems~\cite{ref2,ref3,ref4,ref5,ref6}. Here the cavity-emitter coupling rate is comparable to the decay rate of the system, the rotating-wave approximation (RWA) can be applied to explore the dynamics of the system. With the progress of technology, the ultrastrong-coupling regime has been reached in superconductor and solid-state semiconductor systems~\cite{ref13,ref14,ref15,ref16}, where the light-matter interaction strength can reach the order of $10\%$ of the field frequency or the transition frequency of the quantum emitters~\cite{ref7,ref9,ref10,ref11,ref11-1,ref12}. In the regime, the approach involving the RWA is no longer applicable for describing the dynamics of the system. The counter-rotating terms, corresponding to excitation-number-nonconserving processes, induce virtual transitions between states of the system. Such a regime of cavity QED is not only a new door to study the physics of virtual processes~\cite{ref12-1,ref12-2,ref12-4}, but also has potential applications in modern quantum technology~\cite{ref17,ref18,ref19,ref20,ref20-1}. Recently, a further regime, called the deep-strong-coupling regime with the coupling rate between an oscillator and a two-level system that is greater than or equal to the oscillator frequency, has been presented~\cite{ref21}. For this regime, many novel quantum properties have been discovered, such as the periodic collapse-revival of the system populations during the process of dynamical evolution.

Recently, the ultrafast time-control of the light-matter interaction has been achieved experimentally by applying the intersubband transitions in quantum wells, which lead to unconventional QED phenomena~\cite{ref13}. In addition, the all-optical time switch of strong interaction between a microcavity and a cascade three-level system has been proposed~\cite{ref22,ref23}. A control pulse with a specific arrival time can switch off or on the vacuum Rabi oscillations of the system due to the strong population inversion between the levels. If switch-off fails, the time-control pulse might be applied to destroy suddenly the first-order coherence of the cavity photons. The loss of the coherence can be recreated after the arrival of an additional control pulse. In the process, the time evolution of cavity photon population and the coupling between cavity field and emitter are not affected by the pulses. Such a scheme of control over the light-matter interaction and coherence of the system can be used to perform which-path and quantum-eraser operations, and is of great importance in quantum science and modern optics. In the deep-strong coupling regime, we are now interested to determine whether the all-optical time-control pulse can be used to manipulate the dynamical evolution of the system.

Motivated by the above question, we study in this paper the quantum dynamics of a three-level system interacting with a single-mode resonator in the deep-strong-coupling regime. Here the superconducting quantum circuit with a weakly-anharmonic multilevel structure is considered as the three-level emitter~\cite{ref23-1,ref23-2,ref23-3}. In this case, the system Hamiltonian satisfies $Z^2$ symmetry due to the presence of the counter-rotating terms. So the state space splits into two independent parity chains~\cite{ref21,ref23-3,ref24,ref25}. The dynamical evolution of the system can only occur in a certain parity chain decided by the initial state, and the simultaneous evolution in two parity chains is forbidden. In this process of the time evolution, the photon population and the probability of initial state periodically collapse and revive when the system is originally in a product state or symmetry superposition state. This process happens here and arises from the transitions between states induced by rotating terms and counter-rotating terms.

Interestingly, we find that the dynamical evolution of the system can be controlled by feeding a Gaussian pulse into the system to couple to the three-level emitter. A control pulse with a specific arrival time can suddenly switch off the periodic evolution of the photon population and the collapse-revival of the initial state when the system is originally in a symmetry superposition state. This is because the control pulse breaks the symmetry of the system state, giving rise to destructive quantum interference between different transition paths. This is no longer the case for the asymmetry superposition state and product state. Moreover, the disappearing process of evolution is not irreversible, the evolutions of system populations in a certain parity chain can be recovered instantaneously after the arrival of the second control pulse. This all-optical switch of dynamical evolution of the system in the deep-strong-coupling regime is not only fundamentally intriguing, but also have potential in quantum science and quantum engineering.

Our paper is organized as follows. In Sec.~II, we introduce the model and the system dynamics
in two unconnected parity chains caused by the $Z^2$ symmetry of the Hamiltonian. In Sec.~III, we
show the dynamical evolution of the system without feeding coherent pulses into the system. In Sec.~IV, we demonstrate a scheme for switching off and on the evolution of the system populations. In Sec.~V, we give conclusions of our work.
\section{Model}
\begin{figure}
  \centering
  \includegraphics[width=8.5cm]{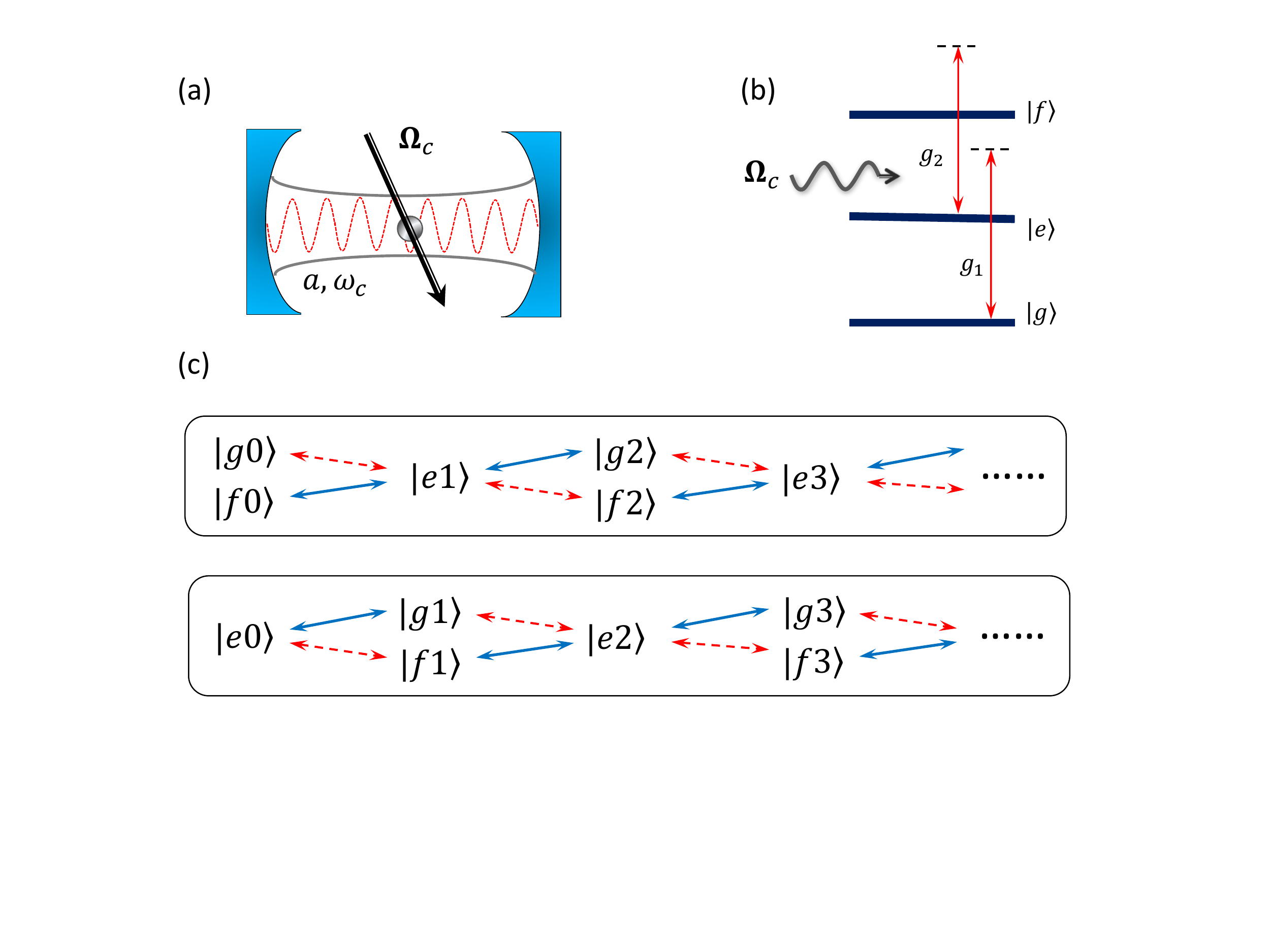}\\
  \caption{(a) Schematics of the system and (b) of the three-level emitter. $g_1$ is the coupling rate between the resonator and the two lowest levels \{$|g\rangle, |e\rangle$\}, $g_2$ is the coupling rate between the resonator and the two excited levels \{$|e\rangle, |f\rangle$\}. $\Omega_c$ denotes the coherent control pulse. (c) Two parity chains describing the dynamical evolution of system decided by the $Z^2$ symmetry, the upper and lower chains correspond to even parity and odd parity, respectively. The transition between two states in the same chain may be connected via either rotating (solid blue arrows) or counter-rotating (dashed red arrows) terms of the Hamiltonian. The first and second indices in state $|A; B\rangle$ denote qubits and resonator, respectively.}\label{fig1}
\end{figure}
As shown in Fig.~\ref{fig1} (a), we consider a quantum system that consists of a single-mode resonator coupled to a cascade three-level emitter in the deep-strong-coupling regime. The superconducting quantum circuit with the weak anharmonicity is considered as the three-level quantum emitter. The coherent control pulse $\Omega_c$ can be feeded into the system to coupled resonantly to two excited states of the emitter. The Hamiltonian of the system is given by ($\hbar=1$)
\begin{equation}\label{eq1}
H=H_0+H_d,
\end{equation}
with
\begin{align}\label{eq2}
H_0=&\omega_c a^\dag a+2\omega_q|f\rangle\langle f|+\omega_q|e\rangle\langle e|+g_1(a^\dag+a)(|e\rangle\langle g|\nonumber\\
&+|g\rangle\langle e|)+g_2(a^\dag+a)(|f\rangle\langle e|+|e\rangle\langle f|),
\end{align}
where the first, second and third terms of Eq. (\ref{eq2}) represent the free Hamiltonian of the single-mode resonator and three-level emitter, and the last two terms of Eq. (\ref{eq2})
 represent the interaction between the resonator and emitter. $H_d$ describes the coupling of coherent a control pulse with two excited states of an emitter with
\begin{equation}\label{eq3}
H_d=\Omega_c(t) |f\rangle\langle e|+\Omega_c^\ast(t)|e\rangle\langle f|.
\end{equation}
Here, $a$ (or $a^\dag$) is the annihilation (or creation) operator of the resonator with resonance frequency $\omega_c$. $\omega_q$ is the transition frequency between adjacent energy levels $|e\rangle$ (or $|f\rangle$) and $|g\rangle$ (or $|e\rangle$) of the emitter, where the level $|g\rangle$ should be considered as the ground state of the emitter.
$\Omega_c(t)=A\cos(\omega t)\exp[-(t-t_c)^2(2\tau^2)]/(\tau\sqrt{2\pi})$ represents the Gaussian control pulse, $A$ and $\tau$ are the amplitude and standard deviation of the pulse, respectively, and $t_c$ is the arrival time of pulse. Note that the Hamiltonian $H_0$ contains the counter-rotating terms of forms $a^\dag |e\rangle\langle g|$, $a |g\rangle\langle e|$, $a^\dag |f\rangle\langle e|$ and $a |e\rangle\langle f|$, which cannot be ignored in the deep-strong-coupling regime.
\begin{figure}
  \centering
  \includegraphics[width=8.5cm]{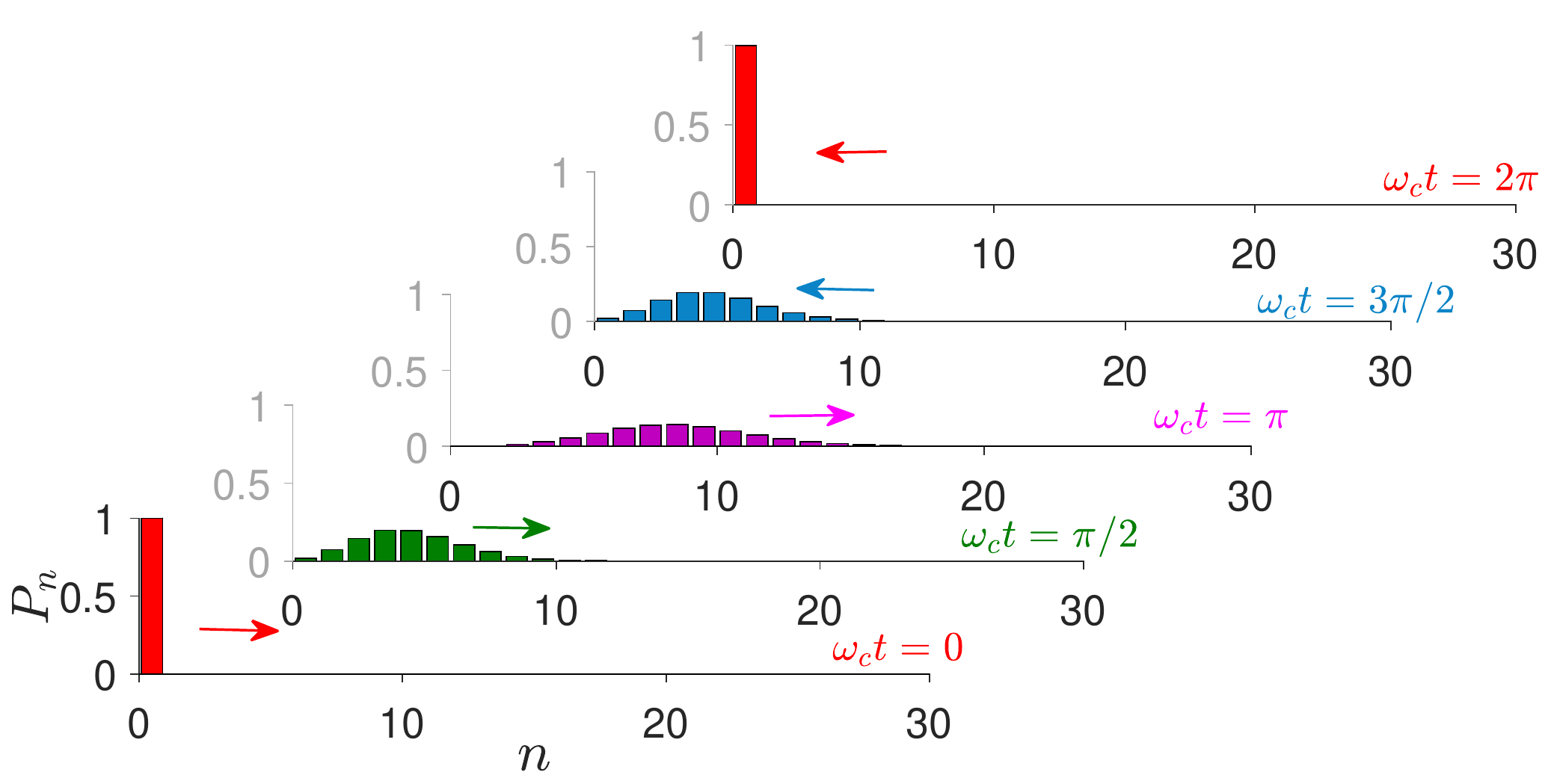}\\
  \caption{Round trip of the photon number wave packet due to the dynamical evolution of the system in the even-parity chain, where the initial state $|\psi_0\rangle=|+0\rangle=(|g0\rangle+|f0\rangle)/\sqrt{2}$, $\omega_q/\omega_c=0$ and $g_1/\omega_c=g_2/\omega_c=1$.}\label{fig2}
\end{figure}

For the system without a coherent control pulse fed into the emitter, the resonator-emitter Hamiltonian $H_0$ has parity (or $Z^2$) symmetry with a well-defined parity operator
\begin{align}\label{eq4}
\Pi=&\exp[i\pi N]=\exp[i\pi(a^\dag a+2|f\rangle\langle f|+|e\rangle\langle e|)]\nonumber\\
=&(|f\rangle\langle f|-|e\rangle\langle e|+|g\rangle\langle g|)(-1)^{a^\dag a},
\end{align}
which commutes with the Hamiltonian $H_0$. $\Pi|p\rangle=p|p\rangle$ ($p=\pm1$), measuring the odd-even parity of the system. The system dynamics moved inside the Hilbert space is split into two unconnected parity chains, i.e., odd parity ($p=-1$) and even parity ($p=+1$)~\cite{ref21,ref23-3,ref24,ref25}, as displayed in Fig.~\ref{fig1}(c). States within each parity chain may be connected via either rotating or counter-rotating terms. For example, in the even-parity chain ($p=+1$), the counter-rotating term $a^\dag |f\rangle\langle e|$ induces the transition $|e1\rangle\rightarrow|f2\rangle$, and the rotating term $a|f\rangle\langle e|$ induces the transition $|e1\rangle\rightarrow|f0\rangle$. These transitions induced by rotating and counter-rotating terms can only occur in a certain parity chain, and the transition between the two parity chains is forbidden due to the protection of $Z^2$ symmetry. Here the counter-rotating terms correspond to the virtual transitions, which are crucial for the dynamical evolution of the system in each parity chain.

\section{Dynamics of system without coherent pulses}
\begin{figure}
  \centering
  \includegraphics[width=8.5cm]{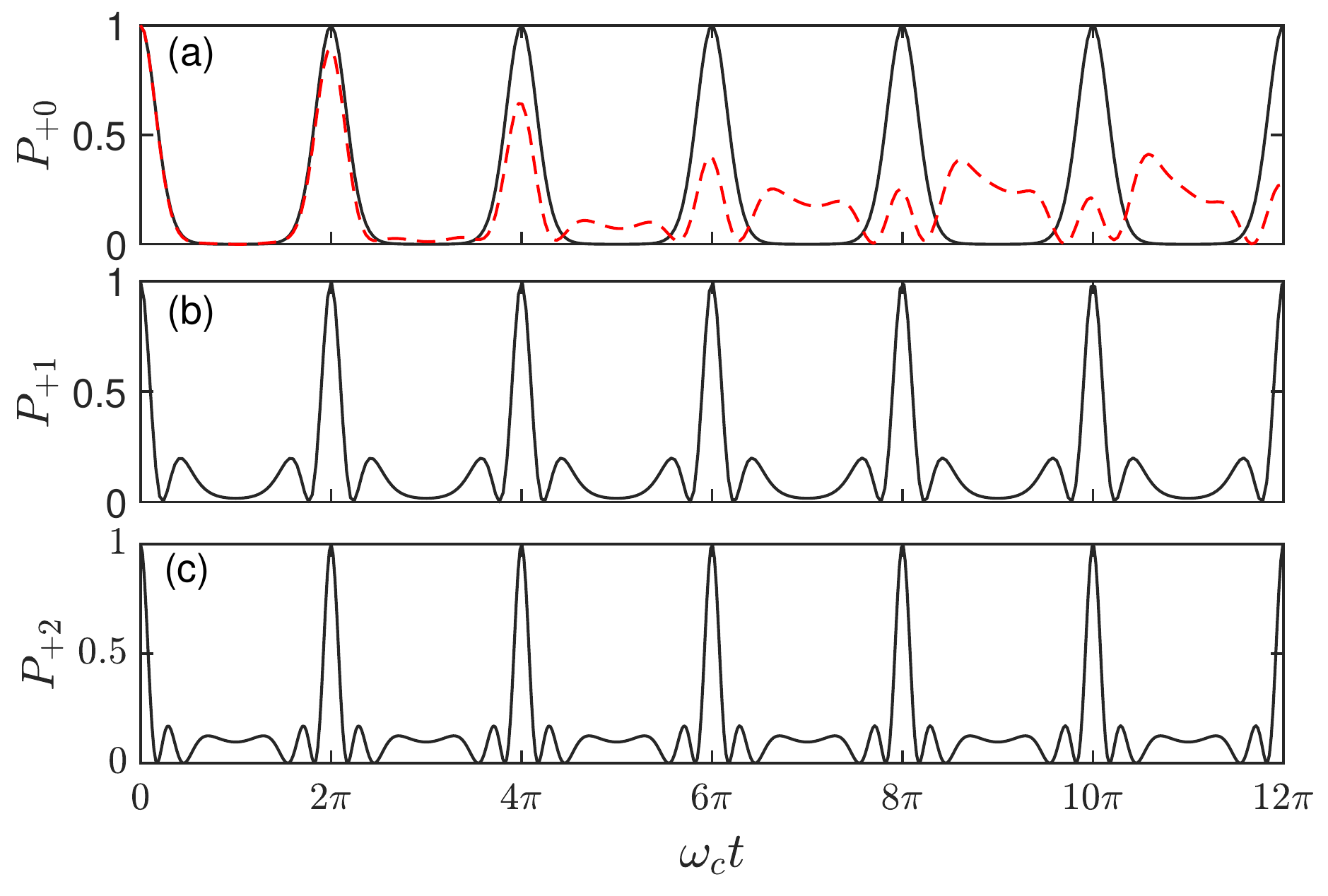}\\
  \caption{The probabilities of initial states at different times of the evolution when $g_1/\omega_c=g_2/\omega_c=1$. Here the initial states (a) $|\psi(0)\rangle=|+0\rangle=(|g0\rangle+|f0\rangle)/\sqrt{2}$, (b) $|\psi(0)\rangle=|+1\rangle=|e1\rangle$ and (c) $|\psi(0)\rangle=|+2\rangle=(|g2\rangle+|f2\rangle)/\sqrt{2}$. The black solid lines and red dashed line correspond to $\omega_q/\omega_c=0$ and $\omega_q/\omega_c=0.2$, respectively.}\label{fig3}
\end{figure}
We consider that our system is designed to operate in the condition of no feeding of the coherent pulses. Assuming  the limit of $\omega_q/\omega_c=0$, the three emitter levels of emitter are degenerate. Here the three-level system still allows only two transitions, i.e., $|g\rangle\leftrightarrow|e\rangle$ and $|e\rangle\leftrightarrow|f\rangle$. The transition between the levels $|g\rangle$ and $|f\rangle$ is forbidden due to the fact that they have the same parities in the emitter~\cite{ref23-1}. In the cases of $g_1/\omega_c=g_2/\omega_c=1$ and the initial state $|\psi(0)\rangle=|+0\rangle=(|g0\rangle+|f0\rangle)/\sqrt{2}$, we study the quantum dynamics of the system
\begin{equation}\label{eq5}
|\psi(t)\rangle=U(t)|\psi(0)\rangle=e^{-iHt}|\psi(0)\rangle,
\end{equation}
where the first and second indices
in state $|+0\rangle$ denote the parity and photon number state, respectively. We find that the photon population $P_n$ goes back and forth on the even chain independently with a period of $T=2\pi/\omega_c$. The round trip of the photon number wave packet is displayed in Fig.~\ref{fig2}. To understand the dynamics of the system more clearly, we calculate numerically the probability of the initial state in the process of system evolution, i.e., $P_{+0}=|\langle\psi(t)|\psi(0)\rangle|^2$, in Fig.~\ref{fig3}(a), which shows the collapses and revivals with a period of $T=2\pi/\omega_c$. The reason for this behavior is that, in the deep-strong-coupling regime, the system allows the occurrence of high-order transition processes induced by the rotating and counter-rotating terms of the Hamiltonian $H_0$. Note that the round trip of the photon number wave packet and the collapse-revival of the initial-state probability cannot be seen in the case of including the RWA. Moreover, we show the time evolutions of the probabilities of the initial states $|\psi(0)\rangle=|+1\rangle=|e1\rangle$ and $|\psi(0)\rangle=|+2\rangle=(|g2\rangle+|f2\rangle)/\sqrt{2}$ in Figs.~\ref{fig3}(b) and ~\ref{fig3}(c), respectively. A series of subpeaks can be seen in the two figures because the dynamics of the system leads to the generation of counterpropagating photon number wave packets, whose spreading in both directions bounce back and forth giving rise to the interference. Comparing Figs.~\ref{fig3}(b) and ~\ref{fig3}(c), we find that the latter curve has more subpeaks than the former one; this result implies that the degree of interference depends on the selection of initial state.

\begin{figure}
  \centering
  \includegraphics[width=8cm]{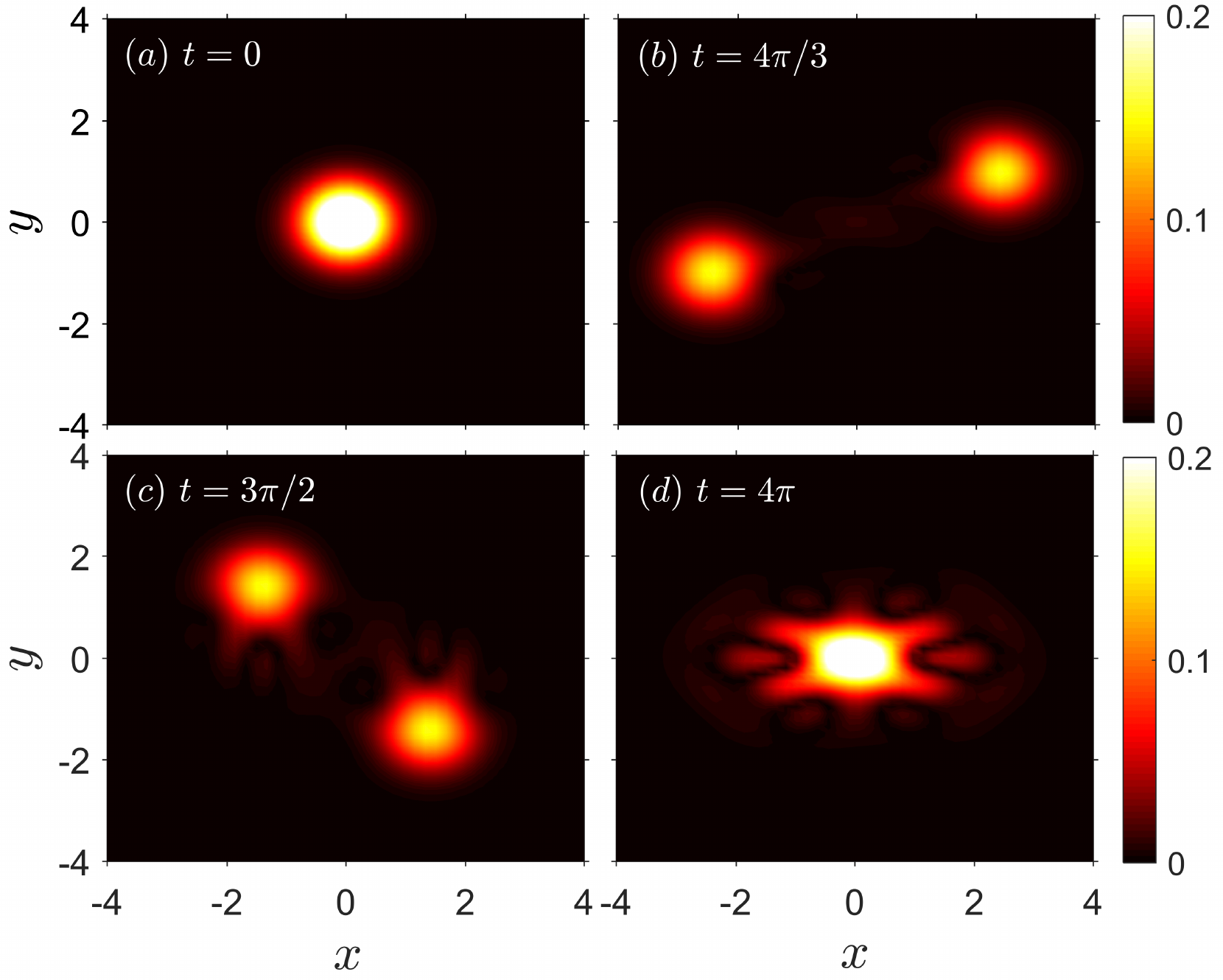}\\
  \caption{The Wigner function $W(x,y)$ of the state for different times of dynamical evolution. The quadrature variables are $x=(a^\dag+a)/\sqrt{2}$ and $y=(ia^\dag-ia)/\sqrt{2}$. For all panels, the initial states $|\psi(0)\rangle=|+0\rangle=(|g0\rangle+|f0\rangle)/\sqrt{2}$, $\omega_q/\omega_c=0.2$ and $g_1/\omega_c=g_2/\omega_c=1$.}\label{fig4}
\end{figure}

Furthermore, we investigate the dynamics of the system in the case of $\omega_q/\omega_c\neq0$, where the degeneracy of the emitter is broken. In Fig.~\ref{fig3}(a), although the dynamical evolution is still in a certain parity chain, the probability of the initial state cannot be revived fully due to the self-interference of the photon number wave packet~\cite{ref21}. To better understand the collapses and revivals that happened here, we present the Wigner function of the state in the phase space, as shown in Fig.~\ref{fig4}.
Comparing Figs.~\ref{fig4}(a) with ~\ref{fig4}(d), we find that the state cannot be revived fully after two periods of evolution, which is consistent with the previous result. From Figs.~\ref{fig4}(b) and ~\ref{fig4}(c), we observe that the initial state is split into two parts to evolve in the even-parity chain. The reason is that the three-level structure of the emitter determines the evolution of initial state with different types of transitions, see Fig.~\ref{fig1}(c).
\section{Switching of the dynamics with coherent pulses}
\begin{figure}
  \centering
  \includegraphics[width=8.0cm]{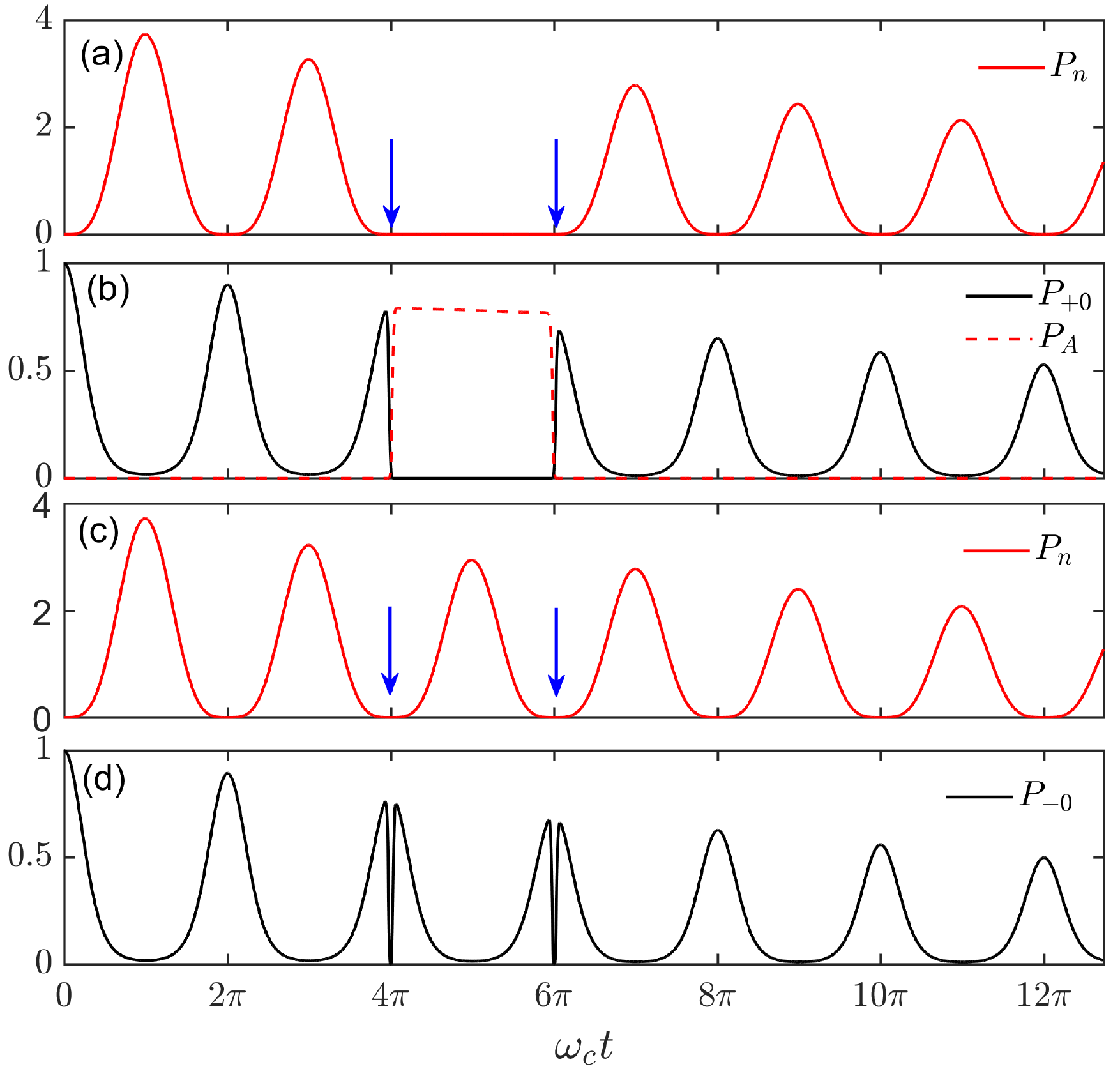}\\
  \caption{(a) (c) Dynamics of the photon population $P_n$ in the presence of two $\pi$ control pulses (blue arrows). (b) Time evolution of the probabilities of the states $|\psi(0)\rangle=|+0\rangle=(|g0\rangle+|f0\rangle)/\sqrt{2}$ and $|\psi_A\rangle=(|g0\rangle-|f0\rangle)/\sqrt{2}$. (d) Time evolution of the probability of the state $|\psi(0)\rangle=|-0\rangle=|e0\rangle$. Here the initial states $|\psi(0)\rangle=|+0\rangle=(|g0\rangle+|f0\rangle)/\sqrt{2}$ for panels (a) and (b), $|\psi(0)\rangle=|-0\rangle=|e0\rangle$ for panels (c) and (d). The system parameters used here are: $\omega_q/\omega_c=\omega/\omega_c=0.01$, $g_1/\omega_c=g_2/\omega_c=0.5$, $\gamma/\omega_c=\kappa/\omega_c=0.005$, $A=\pi$ and $\tau=0.1/\omega_c$.}\label{fig5}
\end{figure}

In this section, we study how to control the time evolution of the system populations by feeding coherent control pulses into the system. To describe the dynamics of the system more realistically, the losses of resonator and quantum emitter need to be taken into account. The effective Hamiltonian of the system reads
\begin{equation}\label{eq6}
H_{\texttt{eff}}=H_0+H_d-i\frac{\kappa}{2}a^\dag a-i\frac{\gamma}{2}(|f\rangle\langle f|+|e\rangle\langle e|),
\end{equation}
where $\kappa$ is the decay rate of the resonator, and $\gamma$ is the relaxation rate of the levels $|f\rangle$ and $|e\rangle$. Starting from the effective Hamiltonian, we derive the evolution of the system
\begin{equation}\label{eq7}
|\psi(t)\rangle=U(t)|\psi(0)\rangle=e^{-iH_{\texttt{eff}}t}|\psi(0)\rangle.
\end{equation}
If $\omega_q/\omega_c=0.01$, i.e., $\omega_q\ll\omega_c$, and the initial state $|\psi(0)\rangle=|+0\rangle=(|g0\rangle+|f0\rangle)/\sqrt{2}$, we calculate the dynamics of the photon population and the probability of the initial state. The results are shown in Figs.~\ref{fig5}(a) and ~\ref{fig5}(b). Note that the initial state is a symmetry even-parity state, whose dynamical evolution can only occur in the even-party chain. At the initial time, the emitter is in the superposition state and the resonator is in the vacuum state, and the average photon number is zero. Then the photon population and the probability of the initial state exhibit periodic collapses and revivals due to the dynamical evolution of system. However, the occupation of the resonator in the nonzero photon state cannot be recovered when the first $\pi$ pulse with the width $\tau=0.1/\omega_c$ arrives in the system at $t_c=4\pi/\omega_c $, while the probability of the initial state suddenly goes to zero. Here the system has come back to the initial state $(|g0\rangle+|f0\rangle)/\sqrt{2}$ before the arrival of the pulse. The results imply that the time evolution of the photon population in the parity chain have stopped. The reason for the sudden disappearance of evolution is that the first $\pi$ control pulse sent at a minimum photon population induces the transitions between levels of quantum emitter $|f\rangle\rightarrow|e\rangle\rightarrow-|f\rangle$, the transitions allow the system to change from a symmetry state $|\psi(t)\rangle=(|g0\rangle+|f0\rangle)/\sqrt{2}$ to an asymmetry state $|\psi\rangle_A=(|g0\rangle-|f0\rangle)/\sqrt{2}$. Note that the width of the pulses may affect the evolution of the system, but here we use the picosecond pulses to excite the emitter~\cite{ref13,ref26}. The width $\tau$ of the pulses is much smaller than the typical time scale $\varsigma$ of the system evolution, where $\varsigma\sim0.02~\texttt{ns}$ ($\omega_c\sim50~\texttt{GHz}$). Thus the evolution of the system is not influenced by the width of the pulses.

 In Fig.~\ref{fig5}(b), the dashed red line represents the probability of the state $|\psi\rangle_A$, which reaches the maximum value when the first pulse arrives in the system. Note that if the system is in an asymmetry state, such as $(|g0\rangle-|f0\rangle)/\sqrt{2}$ and $(|g1\rangle-|f1\rangle)/\sqrt{2}$, the dynamical evolution of the system cannot be carried out in each parity chain. This is because the different transition paths between the two states lead to the destructive quantum interference, making the effective coupling rates between two transition states go to zero, i.e.,
\begin{align}\label{eq8}
\Omega_{\texttt{eff}}=&\langle e1|H_0|\psi_A\rangle\nonumber\\
=&\frac{1}{\sqrt{2}}\langle e1|H_0|g0\rangle-\frac{1}{\sqrt{2}}\langle e1|H_0|f0\rangle=0,~~~~(P=+1)
\end{align}
\begin{align}\label{eq9}
\texttt{or}~~\Omega_{\texttt{eff}}=&\langle e2|H_0|\psi_A\rangle\nonumber\\
=&\frac{1}{\sqrt{2}}\langle e2|H_0|g1\rangle-\frac{1}{\sqrt{2}}\langle e2|H_0|f1\rangle=0.~~~~(P=-1)
\end{align}
\begin{figure}
  \centering
  \includegraphics[width=8.5cm]{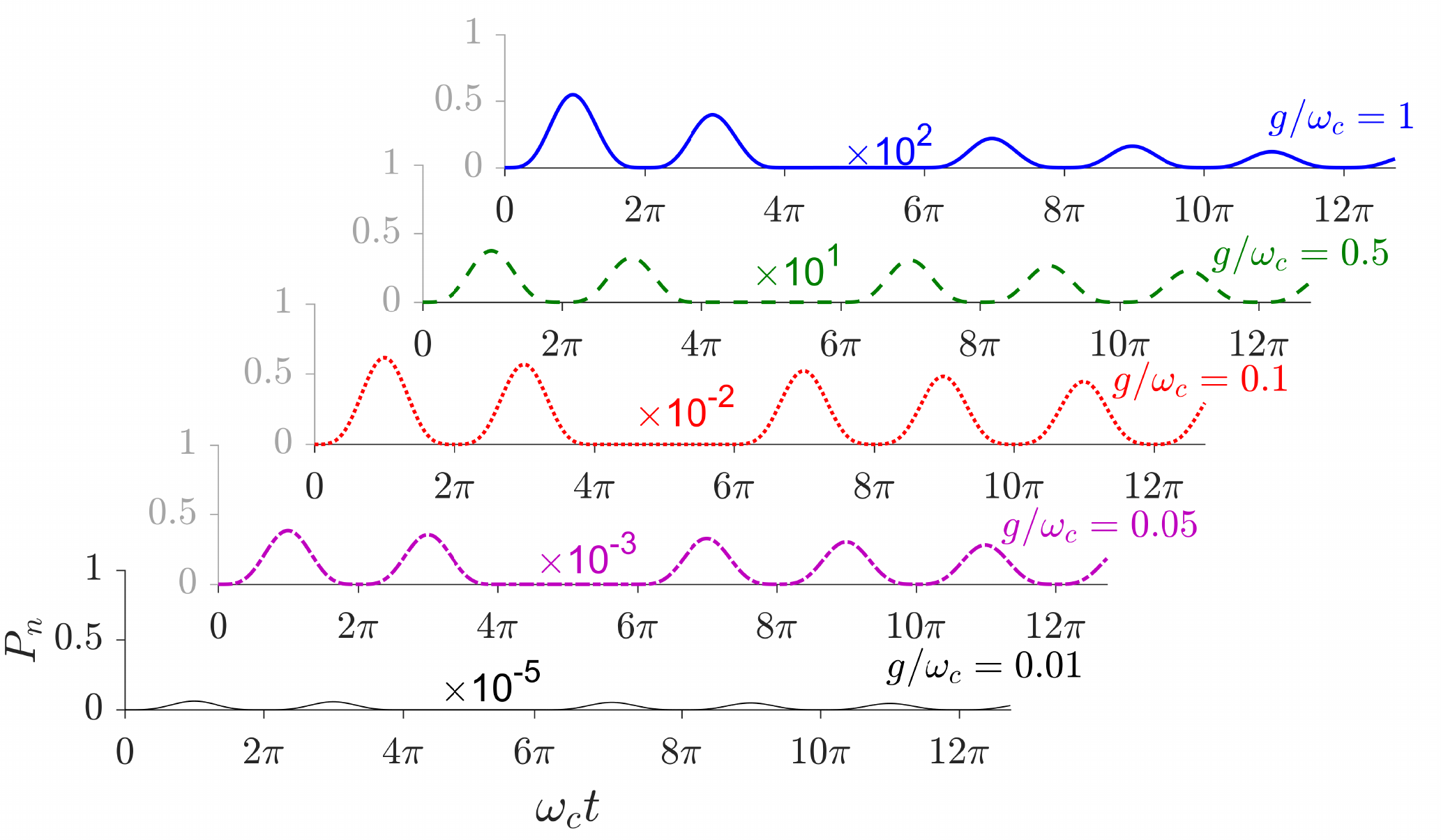}\\
  \caption{Dynamics of photon population
$P_n$ for different coupling rates $g/\omega_c$ in the presence of two $\pi$ control pulses. The initial state $|\psi(0)\rangle=|+0\rangle=(|g0\rangle+|f0\rangle)/\sqrt{2}$ and other system parameters used here are in the same as Fig.~\ref{fig5}.}\label{fig6}
\end{figure}
From Fig.~\ref{fig5}(a), we also find that the average photon number begins to increase when the second control $\pi$ pulse arrives in the system at $t_c=6\pi/\omega_c$. The reason is that the second control pulse, feeded into the system at this time, induces the transitions $|f\rangle\rightarrow|e\rangle\rightarrow-|f\rangle$, which changes the system from an asymmetric state to a symmetric state, i.e., $(|g0\rangle-|f0\rangle)/\sqrt{2}\rightarrow(|g0\rangle+|f0\rangle)/\sqrt{2}$. The result is also displayed in Fig.~\ref{fig5}(b), the sudden recover of the symmetry state induces the rebirth of the dynamical evolution of the system in the even-parity chain. Thus we can switch off and on the evolution of system by feeding the control pulses with specific arrival times into the emitter.
Moreover, in Fig.~\ref{fig6}, we show the dynamics of photon population for different  resonator-emitter coupling strengths $g/\omega_c$. Upon enhancing the coupling strength, the maximum value of photon population increases, since a large value of $g/\omega_c$ allows the presence of higher-order transitions in the parity chain. But the collapses and revivals of the photon population are independent of the changes of coupling strength $g/\omega_c$.

Furthermore, we investigate the quantum dynamics of the system when the initial state is the product state $|e0\rangle$, i.e. $|-0\rangle$. We observe the collapses and revivals of the photon population with a period $2\pi/\omega_c$ in Fig.~\ref{fig5}(c). The time evolution of photon population is not affected by the control pulses with the special arrival times. The result arises from the fact that the first pulse, arrived at a minimum photon population, induces the transitions $|e\rangle\rightarrow-|f\rangle\rightarrow-|e\rangle$ in a very short period of time. These transitions between two excited levels of the emitter cannot change the state of the system. To better understand the transitions induced by the control pulse, we display the time evolution of the probability of the initial state in Fig.~\ref{fig5}(d). The probability shows the collapses and revivals in the first two cycles, and the system changes suddenly from the state $|e0\rangle$ to the state$-|f0\rangle$ when the first control pulse arrives in the system at $t_c=4\pi/\omega_c$. Then the state quickly undergoes the transition $-|f0\rangle\rightarrow-|e0\rangle$. Here the presence of the intermediate state can be seen by the valley of the curve in Fig.~\ref{fig5}(d). Such rapid transitions processes, i.e., $|e0\rangle\rightarrow-|f0\rangle\rightarrow-|e0\rangle$, cannot modify the state of the system.
Similarly, the second control pulse with the arrival time $t_c=6\pi/\omega_c$ induces the transitions $-|e0\rangle\rightarrow|f0\rangle\rightarrow|e0\rangle$.  Thus the collapses and revivals of the photon population are still not affected. To combine the present and previous results, when only the system is originally in the symmetry superposition state, can the specific arrival times of control pulses switch off and on the time evolution of photon population and initial-state probability.
\section{conclusion}
In summary, we have studied the quantum dynamics of the three-level emitter interacting with the single-mode resonator in the deep-strong-coupling regime. In this case, the method involving the RWA is not applied, and the rotating and counter-rotating terms induced the transitions between states in a certain parity chain. The transition between two parity chains is forbidden due to the presence of $Z^2$ symmetry. We have observed the collapses and revivals of the system populations when the system is originally in a product state or symmetry superposition state (a superposition state is formed of state components belonging to the same-parity chain).

Specifically, we have demonstrated that the collapses and revivals of the photon population can be controlled by feeding the control pulses with specific arrival times into the emitter. Such an operation of switching off and on the dynamical evolutions of photon population and initial state probability can be achieved only when the initial state of the system is the symmetry superposition state, whereas this is no longer the case for the asymmetry superposition state and product state. Moreover, we have found that the change of the coupling strength between resonator and emitter does not affect the collapses and revivals of the photon population. Thus the control pulse may be exploited to control the the evolutions of system populations even when the system is in a weaker-coupling regime. This work offers a promising approach to manipulate the dynamical evolution of system populations, which has potential applications in quantum engineering and quantum science.
\section{acknowledgment}

This work is supported by the National Key Research and Development Program of China grant
2016YFA0301203, the National Science Foundation of China (Grant Nos. 11374116, 11574104 and 11375067).

\end{document}